\documentclass[twocolumn,floatfix,aps,showpacs]{revtex4}
\usepackage{amssymb}
\usepackage{amsmath}
\usepackage{graphicx}

\renewcommand{\vec}[1]{\mathbf{#1}}

\begin{document}

\title{Effect of noise on geometric logic gates for quantum computation}
\author{A. Blais}
\email[Email: ]{ablais@physique.usherb.ca}
\author{A.-M. S. Tremblay}
\email[Email: ]{tremblay@physique.usherb.ca}
\affiliation{D\'epartement de Physique and Centre de Recherche sur les Propri\'et\'es
\'Electroniques de Mat\'eriaux Avanc\'es, Universit\'e de Sherbrooke,
Sherbrooke, Qu\'ebec, J1K 2R1, Canada}
\date{\today}
\pacs{03.67.-a,03.65.Vf,74.50.+r}

\begin{abstract}
We introduce the non-adiabatic, or Aharonov-Anandan, geometric phase as a
tool for quantum computation and show how that phase on one qubit can be
monitored by a second qubit without any dynamical contribution. We also
discuss how that geometric phase could be implemented with superconducting
charge qubits. While the non-adiabatic geometric phase may circumvent many
of the drawbacks related to the adiabatic (Berry) version of geometric
gates, we show that the effect of fluctuations of the control parameters on
non-adiabatic phase gates is more severe than for the standard dynamic
gates. Similarly, fluctuations also affect to a greater extent quantum gates
that use the Berry phase instead of the dynamic phase.
\end{abstract}

\maketitle

\section{Introduction}

To be useful, quantum computers will require long coherence time and low
error rate. To attain this goal, good design and careful choice of the
qubit's operation point are crucial~\cite{devoret:2002}. It is however
believed that this will not be enough and that some kind of `software'
protection will be necessary. To achieve this, different strategies have
been suggested: quantum error correction~\cite{steane:99}, decoherence-free
subspaces~\cite{zanardi:97,lidar:98} and bang-bang control~\cite{viola:98}.

Another approach to minimize the effect of imperfections on the controlled
evolution of qubits is to use geometric phases and, in particular, the
adiabatic geometric phase (or Berry's phase)~\cite{berry:84}. Contrary to
the dynamic phase, Berry's phase does not depend on time but is related to
the area enclosed by the system's parameters over a cyclic evolution in
parameter space. It is therefore purely geometric in nature. As a result, it
does not depend on the details of the motion along the path in parameter
space: as long as the area is left unchanged the phase is left unchanged by
imperfections on the path. This tolerance to area preserving imperfections
has suggested to some authors that Berry's phase could be a useful tool for
intrinsically fault-tolerant quantum computation. For example, from the
above argument, one is led to think that Berry's phase gates will not be
very sensitive to random noise along the path~\cite{jones:2000}. Proposals
for the observation and use of this phase for quantum computation have been
given for different physical systems~\cite{jones:2000,ekert:2000,falci:2000}%
. Application of the non-abelian geometric phase~\cite{wilczek:84} to
quantum computation was also the subject of several publications~\cite%
{zanardi:99,duan:2001,choi:2001,faoro:2002}.

In this paper, we consider another type of geometric phase as a tool for
quantum computation: the non-adiabatic, or Aharonov-Anandan (AA), geometric
phase~\cite{aharonov:87}. As Berry's phase, the AA phase is purely
geometric. It is related to the area enclosed by the state vector in
projective space (see below) during a cyclic evolution. One would therefore
believe that quantum gates based on this geometric phase also have some
built-in tolerance to noise about the path. The use of this gate as a tool
for intrinsically fault-tolerant quantum computation was also recently
suggested in Ref.~\cite{xiangbin:2001}.

In this paper we point out that, when compared to Berry's phase, the AA
phase seems to have many advantages for quantum computation. We also discuss
quite generally how to monitor this global phase on one qubit using a second
qubit. Implementation of the AA phase in a symmetric superconducting charge
qubit~\cite{makhlin:2001} is also discussed. Implementation in other quantum
computer architectures is a simple generalization. The main point of this
paper however is to show that the above arguments concerning tolerance to
noise do not hold. Logical gates based on this phase are in fact \emph{more}
affected by random noise in the control parameters than equivalent dynamic
gates. By studying the effect of random noise on the qubit's control
parameters, we are able to obtain a bound on the value of the phase beyond
which the AA phase gate would be advantageous over its dynamical equivalent.
In this way, we show that the AA phase is never useful in practice. This
result is confirmed numerically for different noise symmetries. Moreover,
using the same analytical and numerical approaches, we point out that
quantum gates based on Berry's phase are also more affected by fluctuations
than their dynamical counterparts.

\section{Adiabatic \textit{vs} non-adiabatic geometric phase gates}

Let us begin by recalling the main ideas related to Berry's phase and see
what are its drawbacks for quantum computation applications. Consider a
system whose Hamiltonian $H(t)$ is controlled by a set of external
parameters $\vec{R}\mathit{(t)}$. Upon varying $\vec{R}\mathit{(t)}$
adiabatically, if the system is initially in an eigenstate of $H$ it will
remain in an eigenstate of the instantaneous Hamiltonian. Moreover, if $H$
is non-degenerate on a closed loop $C$ in parameter space such that $\vec{R}%
\mathrm{(0)}=\vec{R}\mathit{(\tau )}$, the final state will differ only by a
phase factor from the initial state. Berry has shown that this phase factor
has both a dynamic and a geometric contribution, the later depending solely
on the loop $C$ in parameter space~\cite{berry:84}. If the initial state is
a superposition of eigenstates $|\psi _{n}\rangle $ of the Hamiltonian, each
of the eigenstates in the superposition will acquire a Berry phase $|\psi
_{n}(\tau )\rangle =U(\tau )|\psi _{n}(0)\rangle =e^{i\phi _{n}}|\psi
_{n}(0)\rangle $ for some real, eigenstate-dependent, phase $\phi _{n}$~\cite%
{anandan:87b}. These phases will generally have both dynamic and geometric
contributions. This is not a cyclic evolution of the state vector but this
does not lead to any ambiguities since Berry's phase is defined over
parameter space.

It follows from the above that the application of adiabatic geometric phases
to quantum computation has several drawbacks. First, quantum computers will
very likely have a short coherence time. To take full advantage of this
short time, the logic operations should be realized as fast as possible. The
adiabaticity constraint means that Berry's phase gates will be slow, thereby
reducing the effective quality factor of the quantum computer.

Another drawback of the adiabatic phase gate is that during the adiabatic
evolution, both geometric and dynamic phases are acquired. The later is not
tolerant to area preserving noise and must be removed. This could be done
using spin-echo like refocusing schemes which require going over the
adiabatic evolution twice~\cite{jones:2000,ekert:2000,falci:2000}. However,
this further increases the time required to realize a single phase gate and
imperfect operation will cause the dynamic phase not to cancel completely,
thereby introducing errors.

A third difficulty is that adiabatic geometric phases are only possible if
non-trivial loops are available in the space of parameters controlling the
qubit's evolution. In other words, the single-qubit Hamiltonian must be of
the form 
\begin{equation}
H=\frac{1}{2}B_{x}(t)\,\sigma _{x}+\frac{1}{2}B_{y}(t)\,\sigma _{y}+\frac{1}{%
2}B_{z}(t)\,\sigma _{z},
\end{equation}
where control over all three (effective) fields $B_{i}(t)$ is possible. Such
control is not possible in most of the current proposals for solid-state
quantum computer architectures. Control over only two fields, say $B_{x}$
and $B_{z}$, is usually the norm. In this case, all loops in parameter space
are limited to the $x$--$z$ plane and the (relative) Berry phase is limited
to integer multiples of $2\pi $, of no use for computation. Control over
fields in all three directions is possible in NMR where Berry phase gates
have been implemented experimentally~\cite{jones:2000}. More recently, Falci 
\textit{et al.}~\cite{falci:2000} have extended the original superconducting
charge qubit proposal~\cite{makhlin:2001} from a symmetric to an asymmetric
design to allow a non-zero $B_{y}$ and therefore non-trivial closed paths in
parameter space.

This need for external control of many terms in the single-qubit Hamiltonian
means additional constraints, experimental difficulties and sources of noise
and decoherence. This is clearly contrary to the efforts now invested in
reducing quantum computer design complexity using the approach of encoded
universality~\cite{bacon:2001}.

As we shall see, all of the above issues, namely slow evolution, need for
refocusing and control over many effective fields, seem to be resolved when
one considers the non-adiabatic generalization of Berry's phase: the
Aharonov-Anandan (AA) phase.

The latter is introduced by restricting oneself, for a given $H(t)$, to
initial states which satisfy 
\begin{equation}
|\psi (\tau )\rangle =U(\tau )|\psi (0)\rangle =e^{i\phi }|\psi (0)\rangle .
\label{cyclic}
\end{equation}
For non-adiabatic evolutions, these so-called cyclic initial states~\cite%
{moore:91} are generally not eigenstates of the system's Hamiltonian but of
the evolution operator. Aharonov and Anandan~\cite{aharonov:87} have shown
that the total phase $\phi $ acquired by such a cyclic initial state in the
interval $[0,\tau ]$ on which it is cyclic is given by the sum of a dynamic (%
$\hbar =1$), 
\begin{equation}
\delta =-\int_{0}^{\tau }\,dt\,\langle \psi (t)|H(t)|\psi (t)\rangle ,
\label{dynamical}
\end{equation}
and of a geometric contribution, 
\begin{equation}
\beta =\phi -\delta .
\end{equation}
The latter is the AA phase. This result is exact, it does not rest on an
adiabatic approximation \emph{but}, it is restricted to cyclic initial
states, for which Eq.~(\ref{cyclic}) holds.

The AA phase is not associated to a closed loop in parameter space, as in
Berry's case, but rather to a closed loop $C^*$ in projective Hilbert space 
\cite{aharonov:87}. For a (pseudo) spin-1/2, which is the system of interest
for quantum computation, $\beta$ is equal to plus or minus half of the solid
angle enclosed by the Bloch vector $\vec{b}\mathit{(t)}$ on the Bloch
sphere. Recall that the Bloch vector is defined through the density matrix
as 
\begin{equation}
\rho(t) = | \psi(t) \rangle\langle \psi(t) | = \frac{1}{2} \left(\openone + 
\vec{b}\mathit{(t)} \cdot \boldsymbol{\sigma} \right),
\end{equation}
where $\openone$ is the identity matrix and $\boldsymbol{\sigma}$ the vector
of Pauli matrices.

Let us now consider the AA phase as a tool for quantum computation. The
first of the above mentioned issues with the adiabatic phase has already
been solved, as the adiabaticity constraint has been relaxed by choosing
appropriate cyclic initial states which depend on the particular evolution
we are interested in.

The second drawback of the adiabatic phase is solved by choosing evolutions
such that 
\begin{equation}
\langle \psi(t) |H(t)| \psi(t) \rangle = 0  \label{delta_kernel}
\end{equation}
at all times. The dynamic contribution (\ref{dynamical}) is thus zero and
only a geometric AA phase is acquired over $C^*$. For (\ref{delta_kernel})
to be zero at all time, the axis of rotation must always be orthogonal to
the state vector. The corresponding paths are then spherical polygons where
each segment lies along a great circle on the Bloch sphere. It is a clear
advantage of the AA phase for computation that such paths exist since there
is then no need for cancellation of the dynamic phase using refocusing
techniques.

To address the third issue, we restrict our attention to Hamiltonians for
which only two control fields are non-zero, 
\begin{equation}
H=\frac{1}{2}B_{x}(t)\,\sigma _{x}+\frac{1}{2}B_{z}(t)\,\sigma _{z}.
\end{equation}

%%%%%%%%%%%%%%%%%%%%%
% SLICE PATH
%%%%%%%%%%%%%%%%%%%%%

\begin{figure}[tbp]
\centering \includegraphics[width=3.25in]{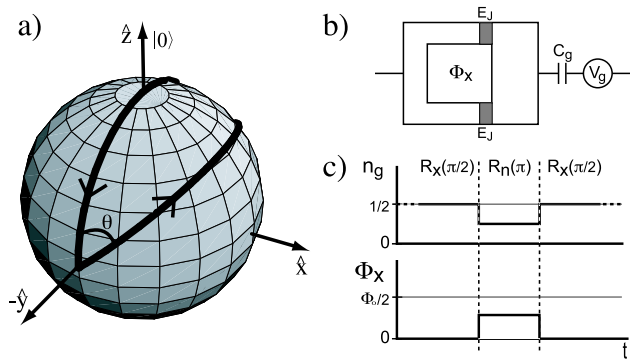}
\caption{a) Evolution of the Bloch vector on the Bloch sphere for the
sequence of pulses (\protect\ref{slice}). The initial (cyclic) state vector
is $| 0 \rangle$. Starting with $| 1 \rangle$ yields a similar path but
centered on the south pole of the Bloch sphere. b) Symmetric charge qubit.
The control parameters are the gate voltage $V_g$ and the external flux $%
\Phi_x$. c) Sequence of external flux $\Phi_x$ and dimensionless gate charge 
$n_g$ implementing $R_z^{AA}(\protect\theta) $. The gate charge is related
to the gate voltage by $n_g=C_gV_g/2e$. Relative amplitude of flux and gate
voltage during $R_n(\protect\pi)$ is used to tune $\protect\theta$, see Fig.~%
\protect\ref{fig_theta}.}
\label{fig_slice}
\end{figure}

If one can turn on and tune the coefficients of $\sigma _{x}$ and $\sigma
_{z}$ simultaneously, the following evolution is possible 
\begin{equation}
R_{z}^{AA}(\theta )\equiv R_{x}(\pi /2)R_{\vec{n}}(\pi )R_{x}(\pi /2),
\label{slice}
\end{equation}
with $\vec{n}=(-\cos \theta ,\mathrm{0},\sin \theta )$ and $B_{n}=\sqrt{%
B_{x}^{2}+B_{z}^{2}}$. This operation acts as $R_{z}^{AA}(\theta
)\:|0\rangle =e^{-i\,\theta }\:|0\rangle $. Figure~\ref{fig_slice}a) is a
plot of this path on the Bloch sphere. Since this path satisfies Eq.(\ref%
{delta_kernel}), the dynamic phase is zero for this evolution and, as a
result, the geometric AA phase is just $-\theta $. By varying the angle of
the axis of rotation $\theta $, it is possible to obtain any geometric
phases. Incidentally, in implementations for which the fields $B_{x}$ and $%
B_{z}$ cannot be non-zero simultaneously, one is restricted to $\vec{n}=\pm 
\vec{z}$ and hence to multiples of $\pi /2$ for $\theta $.

\begin{figure}[tbp]
\centering \includegraphics[width=2.5in]{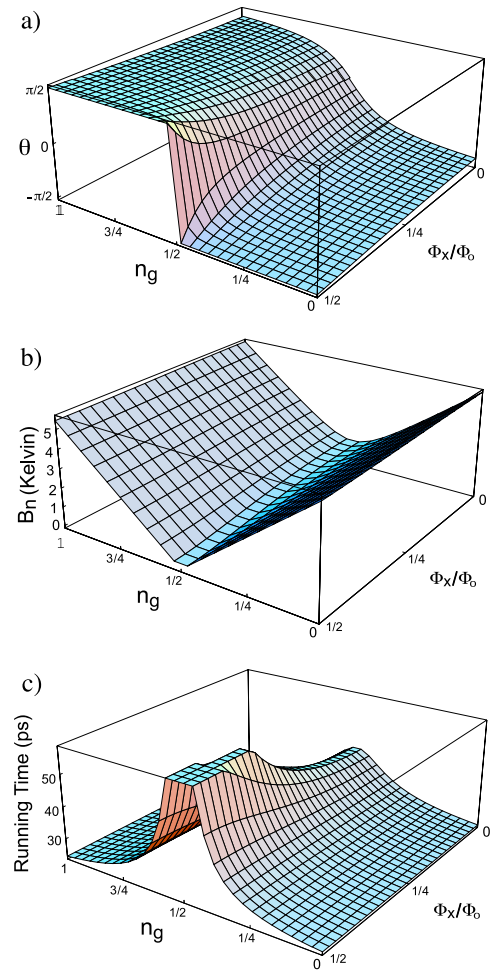}
\caption{a)~Possible values of the geometric phase $\protect\theta =\arctan
[2E_{c}\left( 2n_{g}-1\right) /E_{J}\cos \left( \protect\pi \Phi _{x}/\Phi
_{0}\right) ]$ for the symmetric superconducting charge qubit as a function
of gate charge $n_{g}$ and external flux $\Phi _{x}$ of the rotation $R_{%
\vec{n}}(\protect\pi )$. The characteristic energies of the qubit are chosen
as in Ref.~\protect\cite{nakamura:99}:~$E_{J}=0.6$K and $E_{c}=1.35$K. The
relative phase $2\protect\theta $ can be chosen in the full range $[0,2%
\protect\pi ]$ by an appropriate choice of the control parameters.
b)~Magnitude of the effective field $B_{n}$ as a function of the external
parameters. c)~Total running time of $R_{z}^{AA}(\protect\theta )$ (in
picoseconds) as a function of external control parameters of the $%
R_{n}\left( \protect\pi \right) $ operation in Eq.~(\protect\ref{slice}). We
assume that the $R_{x}\left( \protect\pi /2\right) $ part of the operation
is performed at the fastest possible rate. Due to limitations of voltage and
current (i.e.~flux) pulse generators, actual running time may be larger~%
\protect\cite{nakamura:99}. Finite rise time of the pulses was not taken
into account.}
\label{fig_theta}
\end{figure}

This operation can be implemented, for example, with a symmetric
superconducting charge qubit~\cite{makhlin:2001}, Figure~\ref{fig_slice}b),
by using the sequence of flux and gate voltage of Figure~\ref{fig_slice}c).
This is similar to what was suggested recently in Ref.~\cite{xiangbin:2001}.
Figure~\ref{fig_theta}a) and \ref{fig_theta}b) show respectively the angle $%
\theta$ and the magnitude of the effective field $B_n$ for $R_{\vec{n}}(\pi
) $ as a function of gate voltage and external flux applied on the charge
qubit. Here, $B_z = 4E_c(1-2n_g)$ and $B_x=2E_J\cos(\pi\Phi_x/\Phi_0)$ where 
$\Phi_0=h/2e$ is the flux quantum and $E_c$ and $E_J$ are respectively the
charging and Josephson energies~\cite{makhlin:2001}. Because of the
dependence of $B_n$ on the external parameters, the time $t_n=\pi/B_n$
required to implement $R_{\vec{n}}(\pi )$ depends on the desired geometric
phase $\theta$, Figure~\ref{fig_theta}c).

The gate sequence (\ref{slice}) on the superposition $(a|0\rangle
+b|1\rangle )/\sqrt{2}$ yields 
\begin{equation}
\frac{1}{\sqrt{2}}\left(a\, e^{-i\,\theta }\,|0\rangle +
b\,e^{+i\,\theta}\,|1\rangle\right)  \label{non_cyclic}
\end{equation}
and the phase difference between $|0\rangle $ and $|1\rangle $ has
observable consequences. While this final state depends on the AA phase of
the evolution of $| 0 \rangle$ and $| 1 \rangle$ separately, it is not a
cyclic evolution when acting on their superposition.

For the adiabatic (Berry) phase, a similar situation does not cause any
ambiguities. In that case, as stated earlier, a superposition of eigenstates
does not yield a cyclic evolution for the state vector either. Nevertheless,
the phase acquired by each eigenstate still has a contribution which is
geometric in nature since cyclicity is not required in projective space but
in the Hamiltonian parameter space~\cite{anandan:87b}.

\begin{figure}[tbp]
\centering \includegraphics[width=1.75in]{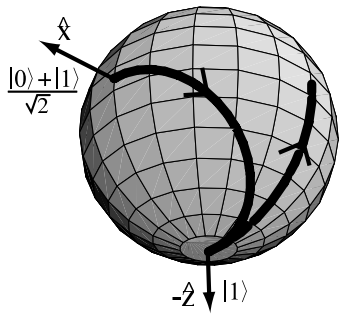}
\caption{The sequence of rotations (\protect\ref{slice}) applied on the
superposition of states $(|0\rangle +|1\rangle)/\protect\sqrt{2}$ does not
yield a closed path on the Bloch sphere.}
\label{fig_open_path}
\end{figure}

In the non-adiabatic case however, there is clearly no closed loop on the
Bloch sphere, as shown on Figure~\ref{fig_open_path}, and identifying the AA
phase according to Aharonov and Anandan's original definition is more
subtle. This situation has suggested to some authors~\cite{bouchiat:88} that
the AA phase is not observable for any evolution on an isolated quantum
system. The reason is that the AA phase is defined only for cyclic
evolutions and, since global phase factors are not physical, observable
properties are unchanged for such evolutions.

%%%%%%%%%%%%%%%%%%%%%
% DIRECT OBSERVATION
%%%%%%%%%%%%%%%%%%%%%

While a non-abelian version of the non-adiabatic phase can be defined and
the phase factors in (\ref{non_cyclic}) can be seen as geometric~\cite%
{anandan:88}, a direct observation of the AA phase as in the NMR experiment
of Suter \textit{et al.}~\cite{suter:88} is interesting but will require
more than one qubit. In the language of quantum computation, the analog of
this NMR experiment is to use a second qubit to `monitor' the phase on the
first one. Explicitly, start with a two-qubit state assuming the first qubit
is in an arbitrary linear superposition 
\begin{equation}
(a|0\rangle +b|1\rangle )\;|0\rangle .  \label{geo_initial}
\end{equation}%
Then, apply the sequence (\ref{slice}) on the second qubit, conditionally on
the first qubit to be $|1\rangle $ 
\begin{equation}
\begin{split}
C_{R_{z}^{AA}}& \equiv C_{\mathrm{NOT}}\,R_{z2}^{AA}(-\theta /2)\,C_{\mathrm{%
NOT}}\,R_{z2}^{AA}(\theta /2) \\
& =%
\begin{pmatrix}
1 &  &  &  \\ 
& 1 &  &  \\ 
&  & e^{-i\theta } &  \\ 
&  &  & e^{+i\theta }%
\end{pmatrix}%
.
\end{split}
\label{evolution}
\end{equation}%
The operation $C_{\mathrm{NOT}}$ is the Controlled-NOT applied on the two
qubits, the first one acting as control. $R_{z2}^{AA}(\pm \theta /2)$ is (%
\ref{slice}) applied on qubit 2 only. This yields 
\begin{equation}
\begin{split}
C_{R_{z}^{AA}}\:(a|00\rangle +b|10\rangle )& =a|00\rangle +be^{-i\,\theta
}|10\rangle \\
& =(a|0\rangle +e^{-i\,\theta }b|1\rangle )\;|0\rangle .
\end{split}
\label{evolution2}
\end{equation}%
The net result is equivalent to a geometric phase gate on the first qubit.
It can be observed from the first qubit by interference~\cite{neutron}.
There is no ambiguity in defining the AA phase in this situation~: The
second qubit undergoes a cyclic evolution and its phase is measurable since
the evolution of the total system is not cyclic.

The controlled-NOT can be realized as 
\begin{eqnarray}
C_{\mathrm{NOT}} &=&e^{-i3\pi /4}\,R_{x2}(3\pi /2)\,C_P(3\pi /2)\,R_{z2}(\pi
/2)  \notag \\
&&R_{x2}(\pi /2)\,R_{z2}(\pi /2)\,R_{z1}(\pi /2)\,C_P(3\pi /2).  \label{cnot}
\end{eqnarray}
This particular sequence is specific to quantum computer implementations
having the control phase shift gate 
\begin{equation}
C_P(\gamma )=e^{-i\gamma \,\sigma _{z}\otimes \sigma _{z}/2}
\end{equation}
in their repertory but similar sequences can be found for other
implementations. For the charge qubit, such a $\sigma _{z}\otimes \sigma
_{z} $ interaction can be implemented by capacitive coupling~\cite%
{falci:2000}.

Using (\ref{slice}) and (\ref{cnot}), it is possible by inspection to
`compile' the total sequence (\ref{evolution}) from $2\times (7+3)=20$ down
to $18$ elementary operations. Moreover, one can verify that the dynamic
phase cancels in (\ref{evolution}). This therefore corresponds to a purely
geometric 2-qubit operation. This logic gate however involves the
application of $18$ elementary gates, a number that is quite large for a
gate whose purpose is to implement a \textquotedblleft
noiseless\textquotedblright\ (geometric) phase-shift gate.

%%%%%%%%%%%%%%%%%%%%%
% IMPERFECTIONS
%%%%%%%%%%%%%%%%%%%%%

\section{Tolerance to noise in control parameters}

A central issue to address in a pragmatic way is tolerance to imperfections.
If non-adiabatic geometric logical gates are to be useful for computation,
there should be some tolerance to fluctuations in the control parameters.
Fluctuations of the control fields will introduce imperfections in the
angles and axes of rotation of the gates implementing the geometric
evolution. These imperfections change the overall unitary evolution applied
on the qubit and the corresponding final phase may now have a dynamic
component. It is important to note that whether the imperfections affect the
dynamic or the geometric component is not relevant for our analysis. Any
unwanted phase factor represents an error on the quantum computation. In the
following, we thus focus on the errors on the total phase coming from
fluctuations in the control parameters around the values that are needed to
achieve the desired unitary transformations in the non-fluctuating case.

Let us consider first the effect of the simplest of such errors: an error $%
\epsilon$ in the angle of the first gate of the sequence~(\ref{slice}) 
\begin{equation}
R_x(\pi/2)R_{\vec{n}}(\pi)R_x(\pi/2+\epsilon).  \label{error_slice}
\end{equation}
We do not consider the extra gates (\ref{evolution}) for the moment.
Evidently, this is not an area preserving error and one should not expect
the AA phase to be invariant in this circumstance. However, this is exactly
the type of errors which will occur if the control field $B_x (t)$ is
fluctuating.

That the non-adiabatic phase gate is not tolerant to this error is easily
checked by applying the erroneous sequence (\ref{error_slice}) on the state $%
|0\rangle $ to obtain 
\begin{equation}
\cos (\epsilon /2)\,e^{-i\,\theta }\,|0\rangle -i\sin (\epsilon
/2)\,e^{+i\theta }\,|1\rangle .
\end{equation}%
The evolution is not cyclic anymore and we cannot define the AA phase in
this situation (at least not in the computational basis). Note that to first
order in $\epsilon $, the non-cyclicity remains and therefore non-adiabatic
phase gates are not tolerant to small imperfections. Small errors can take
the state vector out of great circles and bring in a dynamical contribution.
In worse cases, as above, the evolution is no longer cyclic and the AA phase
can no longer be defined in the computational basis.

It is possible to get a more complete picture of the effect of random noise
on the non-adiabatic phase gate and see how it compares to the simpler
dynamic phase gate 
\begin{equation}
R_z(\theta) = e^{-i\theta\:\sigma_z/2}
\end{equation}
by studying the Hamiltonian 
\begin{equation}
H= \frac{1}{2} \sum_{i=x,z}(B_i(t)+\delta B_i(t))\,\sigma_i.
\end{equation}
Here, $\delta B_i$ represents fluctuations of the control field $B_i$. It is
believed that fluctuations of the control fields are the most damaging
sources of noise and decoherence for solid-state qubits~\cite{makhlin:2001}.
For the charge qubit of Figure~\ref{fig_slice}b), this corresponds to
Nyquist-Johnson noise in the gate voltage $V_g$ and in the current
generating the flux $\Phi_x$.

Without noise, $R_{z}^{AA}(\theta /2)$ and $R_{z}(\theta )$ have the same
effect. To compare these gates in the presence of noise, we simply use the
composition property of the evolution operator 
\begin{equation}
U(t)=\mathcal{T}\:e^{-i\int_{0}^{t}\,dt^{\prime }H(t^{\prime
})}=\lim_{N\rightarrow \infty }\prod_{n=1}^{N}U(n),  \label{semi-group}
\end{equation}
where $U(n)=\exp {(-iH(n)\,t/N)}$ and $H(n)$ is the Hamiltonian during the $%
n^{\mathrm{th}}$ interval. We use units where $\hbar =1.$ To simulate noise,
the fields $\delta B_{i}(n)$ are chosen as independent random variables
drawn from a uniform probability distribution in the interval $\pm \delta
B_{max}$. Without noise, the decomposition (\ref{semi-group}) is of course
exact, whatever the value of $N$, since the logical operations $%
R_{z}^{AA}(\theta /2)$ and $R_{z}(\theta )$ are implemented by piecewise
constant Hamiltonians. With noise, we assume that the $\delta B_{i}$ are
time independent during the interval $\Delta t\equiv t/N_{i}$. We then
define $\Delta t$ as the noise correlation time. It will be assumed to be
the same during the application of any elementary operation $R_{i}.$ With
the decomposition of Eq.~(\ref{semi-group}), the evolution is explicitly
unitary.

To compare the two operations, we compute the trace distance~\cite%
{nielsen-chuang} 
\begin{equation}
D(U,V)=\mathrm{Tr}\left\{ \sqrt{(U-V)^{\dag }(U-V)}\right\}
\end{equation}
with respect to the noiseless $R_{z}(\theta )$ gate. We reached the same
conclusions when the average fidelity~\cite{nielsen:2002} was used
numerically to compare noisy and noiseless gates. The trace distance $D(U,V)$
takes values between 0 and 4, with $D(U,V)=0$ only for $U$ and $V$ equal.
Thus, if the non-adiabatic gate is to be more tolerant to noise than its
dynamic counterpart then 
\begin{equation}
D\boldsymbol{(}\tilde{R}_{z}^{AA}(\theta /2),R_{z}(\theta )\boldsymbol{)} <D%
\boldsymbol{(}\tilde{R}_{z}(\theta ),R_{z}(\theta )\boldsymbol{)}
\label{criterion}
\end{equation}
should hold. The tilde is used here to denote noisy logical gates.

To compute the distance, we expand $U(n)$ in (\ref{semi-group}) to first
order in $\delta B$ and $t/N$ and average the distance obtained from this
approximation by applying the Central Limit Theorem to the variables $%
X_{i}\equiv \sum_{i=1}^{N}\delta B_{i}\left( n\right) $. In addition, we
note that the time necessary to complete $R_{i}\left( \phi \right) $ is $%
t_{i}=N_{i}\Delta t=\phi /B_{i}$. For the geometric gate, this leads to $%
N_{n}B_{n}=2N_{x}B_{x}$ since the rotation angles involved in Eq.~(\ref%
{slice}) are $\pi $ and $\pi /2$ respectively. In this way, we obtain in the
presence of noise along $x$ and $z$ 
\begin{subequations}
\label{D_approx}
\begin{eqnarray}
\langle D\boldsymbol{(}\tilde{R}_{z}^{AA}(\theta /2),R_{z}(\theta )%
\boldsymbol{)}\rangle & \approx& \sqrt{\frac{\pi ^{3}}{12}\left( \frac{1}{%
B_{x}^{2}}+\frac{1}{B_xB_{n}}\right) } \frac{\delta B_{\mathrm{max}}}{\sqrt{%
N_x}};  \notag \\
& & \\
\langle D\boldsymbol{(}\tilde{R}_{z}(\theta ),R_{z}(\theta )\boldsymbol{)}%
\rangle & \approx & \sqrt{\frac{\pi }{6}}\,\frac{\theta \,\delta B_{\mathrm{%
max}}/B_{z}}{\sqrt{N_z}},
\end{eqnarray}
where $B_{x}$, $B_{n}$ and $B_{z}$ are the magnitudes of the effective
fields used to implement respectively $R_{x}(\pi /2)$, $R_{\vec{n}}(\pi )$
and $R_{z}(\theta )$. As $N_{i}$ gets smaller, the noise is constant on a
larger portion of the evolution and excursions on the Bloch sphere farther
away from the original path are possible. The distance between the noisy and
noiseless gates therefore increases as $N_{i}$ diminishes.

\begin{figure}[tbp]
\centering \includegraphics[width=3in]{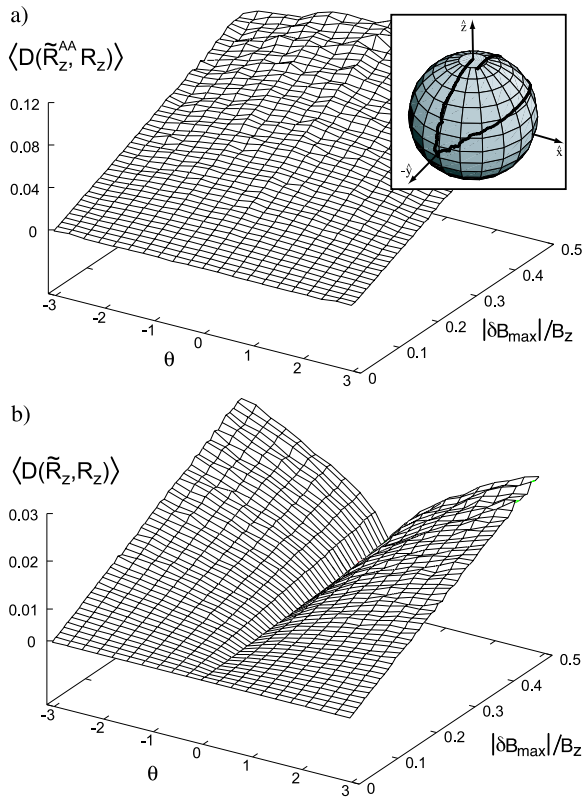}
\caption{\label{fig_distance} Trace distance as a function of $\protect\theta $ and maximum
amplitude of the noise averaged over 600 realizations of the noise. Noise is
along $x$ and $z$ and is in units of the maximal value of the effective
field in the $z$ direction $B_{z}=4E_{c}$. a)~Averaged trace distance
between a noisy AA-phase gate and the corresponding noiseless dynamic $R_{z}$
gate. The inset shows a path with random noise obtained from the numerical
calculation. The path is not closed and the evolution is not cyclic.
b)~Similar to a) but for the noisy dynamic gate $R_{z}$. In both cases, the
noise correlation time is taken as $\Delta t=\hbar /(4E_{c}\protect\gamma )$
with $\protect\gamma =300$. The charging and Josephson energies are taken as
in Fig.~\protect\ref{fig_theta}.}
\end{figure}

Figure \ref{fig_distance} shows a numerical verification of these relations.
The weak dependence of $\langle D\boldsymbol{(}\tilde{R}_{z}^{AA}(\theta
/2),R_{z}(\theta )\boldsymbol{)}\rangle $ on $\theta $ through $B_{n}$ is
apparent in Fig.~\ref{fig_distance}a). For $\langle D\boldsymbol{(}\tilde{R}%
_{z}(\theta ),R_{z}(\theta )\boldsymbol{)}\rangle $, the dependence goes as $%
\sqrt{\theta }$ since $N_{z}\propto \theta $, Fig.~\ref{fig_distance}b). The
agreement between the analytical and numerical results was very good, with
an error of about $3\%$ in both cases. Our first order estimates are then
enough for this level of noise. Systems where the noise is of larger
amplitude will most probably not be relevant for quantum computation so, for
all practical purposes, this approximation should be enough.

Using the analytical estimates (\ref{D_approx}), the criterion (\ref%
{criterion}), and taking the noise correlation time to be equal for dynamic
and geometric gates, we obtain a bound on the angle $\theta $ beyond which
the geometric gate becomes favorable over the dynamic one, 
\end{subequations}
\begin{equation}
\theta _{b}>\pi \left( \frac{B_{z}}{B_{x}}+\frac{B_{z}}{B_{n}}\right) .
\label{BorneAngle}
\end{equation}
Taking $B_{z}/B_{x}\approx B_{z}/B_{n}\approx 1$, we obtain that for $\theta
_{b}\gtrsim 2\pi $ the geometric gate will be less affected by noise than
its dynamic counterpart. For the charge qubit, $B_{z}$ and $B_{x}$ are fixed
respectively by the charging energy $E_{c}$ and Josephson energy $E_{J}$. To
encode efficiently information in the charge degree of freedom, the
inequality $E_{C}\gg E_{J}$ must be satisfied~\cite{makhlin:2001}. The bound
obtained with $B_{z}/B_{x}\approx B_{z}/B_{n}\approx 1$ is therefore a lower
bound on $\theta _{b}$. Since $\theta _{b}>2\pi $, the non-adiabatic
geometric gate is never useful in practice. In particular, with the energies
used in Fig.~\ref{fig_theta}, we obtain $\theta _{b}\gtrsim 2.5\pi $ as a
lower bound. More generally, since the logical states of a qubit are the
eigenstates of $\sigma _{z}$, $B_{z}$ should be larger than $B_{x}$ for the
logical basis to be the `good' basis. We therefore expect this lower bound
to hold for most quantum computer architectures.

We also obtained the analogs of the above results Eqs.~(\ref{D_approx}) and (%
\ref{BorneAngle}) when the noise is along $z$ only and also found the
geometric gate more sensitive to noise than the dynamical one.

The effect of decoherence on the AA phase gate was also studied numerically
by Nazir \textit{et al.} for non-unitary evolutions~\cite{nazir:2002}. They
reach the same conclusion on the sensitivity to noise of the AA phase gate.
Since they can deal with more general noise than we do here, their approach
is more general than ours but is entirely numerical. Our objective here was
to include only the kind of noise to which geometric gates were previously
suggested to be tolerant: unitary random noise about the path.

The approach used here to quantify the effect of fluctuations can be used
for Berry's phase gates as well. We consider the pulse sequence used in the
NMR experiment of Ref.~\cite{jones:2000} and simplified in~\cite{nazir:2002}%
. The system Hamiltonian now takes the form 
\begin{equation}
H=\frac{\Delta }{2}\,\sigma _{z}+\frac{\omega _{1}}{2}\left( \cos \phi
\,\sigma _{x}+\sin \phi \,\sigma _{y}\right) .  \label{H_NMR}
\end{equation}
The sequence of operations used in Ref.~\cite{jones:2000} starts with the
field along the $z$ axis ($\omega _{1}=0$). The parameter $\Delta $ is
assumed fixed throughout. The field is first adiabatically tilted in the $x$%
--$z$ plane by increasing $\omega _{1}$ at $\phi =0$ up to some maximal
value $\omega _{1\,max}$. The field now makes an angle $\theta _{\mathrm{cone%
}}=\arccos (\Delta /\sqrt{\Delta ^{2}+\omega _{1\,max}^{2}})$ with respect
to the $z$ axis. With $\omega _{1}$ kept constant, $\phi $ is then
adiabatically swept from $\phi =0$ to $\phi =2\pi $. To obtain a purely
geometric operation, the dynamic phase is refocused by repeating the above
operations in reverse between a pair of fast $R_y(\pi)$ rotations. The final
relative phase is then purely geometric and has the value $\gamma =4\pi
(1-\cos \theta _{\mathrm{cone}})$~\cite{jones:2000}.

To study the effect of noise for this sequence, we again use the composition
property (\ref{semi-group}) and a Trotter decomposition for (\ref{H_NMR}).
In the same way as above, we then obtain in the case of noise along $x$, $y$%
, and $z$ and assuming that the $R_{y}(\pi )$ rotations are noiseless, 
\begin{equation}
\langle D\boldsymbol{(}\tilde{R}_{z}^{Berry}(\gamma ),R_{z}(\gamma )%
\boldsymbol{)}\rangle \approx \frac{4}{\sqrt{3\pi }}\,\delta B_{\mathrm{max}%
}\,\sqrt{\frac{T_{T}^{2}}{N_{T}}+\frac{T_{\phi }^{2}}{N_{\phi }}},
\end{equation}
where $T_{T}$ is the time taken to tilt the field in the $x$--$z$ plane and $%
T_{\phi }$ the time for the $\phi $ sweep. As in (\ref{D_approx}), the
larger $N_{T}$ and $N_{\phi }$ are, the smaller is the noise correlation
time. Agreement of this result with numerical calculations (not shown) is
excellent. The adiabaticity constraint means that $T_{T}$ and $T_{\phi }$
must be large and therefore that, for all practical purposes, the Berry's
phase gate is worse than its dynamic equivalent. The conclusion is the same
for all the different types of noise tested numerically. For the $\omega_1$
tilt, these are noise along $x$ only and uncorrelated noise along $x$ and $z$%
. For the $\phi$ sweep, we took identical noise along $x$ and $y$, and
tested its effect with and without uncorrelated noise along $z$. Because of
the adiabatic constraint, the Berry's phase gate is also worse than the AA
phase gate. This is the conclusion reached as well in Ref.~\cite{nazir:2002}
in the case of non-unitary evolutions. The possibility~\cite{ekert:2000} to
find a point of operation where conditional phase shifts are insensitive, to
linear order, to noise in $\omega _{1}$ $\left( B_{x}\right) $ may however,
in very special cases, be an advantage of Berry-phase gates for coupled
qubits.

The overall results of this section can be understood intuitively rather
simply. To implement logical gates that use geometric phases (adiabatic or
not), one needs to apply a sequence of unitary transformations that take the
Bloch vector around a closed path. In the presence of noise in the control
fields, that sequence does not take the Bloch vector around a closed path
anymore. Since all that counts is the overall phase of the unitary
transformation, this phase will be more affected in the long sequences of
unitary transformations necessary for geometric gates than in the shorter
sequences necessary for purely dynamical gates. We may point out that if the
noise has a special symmetry that makes it area preserving, then this
symmetry might allow quantum error correction~\cite{steane:99},
decoherence-free subspaces~\cite{zanardi:97,lidar:98} or bang-bang
techniques~\cite{viola:98} to be used with more success than geometric gates.

\section{Conclusion}

In summary, we have considered the AA phase as a tool for quantum
computation. This phase solves many of the problems of Berry's phase gate.
Namely, it can be implemented faster, does not require refocusing of a
dynamic component and involves control over only two effective fields in the
one-qubit Hamiltonian. We showed how the AA phase of one qubit can be
monitored by a second qubit without extra dynamical phase. As an example,
details of the implementation of the AA phase with a symmetric charge qubit
were given. Application of these ideas to other quantum computer
architectures is a simple generalization.

When the effect of noise in the control parameters is taken into account, it
appears that practical implementations of logical gates based on geometric
phase ideas, both adiabatic and non-adiabatic, are more sensitive to noise
than purely dynamic ones, contrary to what was previously claimed. We have
checked how noise affects the overall unitary transformations that, in the
noiseless case, implement purely geometric logical gates. The analytical
results were confirmed numerically and for a wide range of noise symmetries.
This is in agreement with the recent work of Ref.~\cite{nazir:2002}. In the
present work however, we focused our attention on the type of noise to which
the geometric logical gates were previously assumed to be tolerant.

The use of the AA phase for quantum computation purposes therefore seems to
be of little practical interest. It is however of fundamental interest to
observe this phase and a direct observation with the symmetric
superconducting charge qubit seems possible.

\begin{acknowledgments}
We thank S. Lacelle, D. Poulin, H. Touchette and A.M. Zagoskin for helpful
discussions and A. Maassen van den Brink for comments on the manuscript and
useful discussions. This work was partially supported by the Natural
Sciences and Engineering Research Council of Canada (NSERC), the Intelligent
Materials and Systems Institute (IMSI, Sherbrooke), the Fonds pour les
Chercheurs et l'Aide \`{a} la Recherche (FCAR, Qu\'{e}bec), D-Wave Systems
Inc.~(Vancouver), the Canadian Institute for Advanced Research and the Tier
I Canada Research Chair program (A.-M.S.T). Part of this work was done while
A.-M.S.T was at the Institute for Theoretical Physics, Santa Barbara, with
support by the National Science Foundation under grant No. PHY94-07194.
\end{acknowledgments}

%\bibliography{/Users/ablais/Documents/articles/bibliographie/ref.bib}

\begin{thebibliography}{99}
\bibitem{neutron} This is equivalent, for example, to a neutron
interferometry experiment where the first qubit in Eq.(\ref{geo_initial})
represents position and the second spin of the neutron. Since a qubit has
only, by construction, a two-dimensional Hilbert space, it is necessary to
use more then one qubit to mimic interferometry experiments where more than
one degree of freedom of a single particle are used.
\end{thebibliography}
%\end{document}

\end{document}